\documentclass[conference]{IEEEtran}

\usepackage{ifpdf}
\usepackage{subfig}
\usepackage{cite}
\ifCLASSINFOpdf
\usepackage[pdftex]{graphicx}
\else
\fi
\usepackage{amsmath}
\usepackage{amssymb}
\usepackage{algorithmic}
\usepackage{array}
\usepackage{url}
\usepackage{graphicx}
\usepackage{adjustbox,lipsum}
\usepackage{romannum}
\usepackage{siunitx}
\usepackage{gensymb}
\usepackage[english]{babel}
\usepackage[autostyle]{csquotes}
\usepackage{multirow}
\usepackage{tabu}
\usepackage{caption}
\begin{document}

\title{Evaluating PEV's Impact on Long-Term Cost of Grid Assets}

\author{\IEEEauthorblockN{Daijiafan Mao, Danielle Meyer and Jiankang Wang}
\IEEEauthorblockA{Department of Electrical and Computer Engineering\\The Ohio State University\\
Columbus, Ohio 43210\\
mao.156@osu.edu; meyer.758@osu.edu; wang.6536@osu.edu}}

\maketitle
\noindent
\begin{abstract}
With increasing penetration and improving fast charging technologies, Plug-in Electric Vehicles (PEV) exert a disruptive influence on power delivery systems. The impulsive and high-power-density characteristics of PEV make conventional assessment methods of load impact unsuitable. This paper proposes an integrated method to investigate the long-term impact of PEV charging on temporal response and depreciation of grid assets in sub-transmission and distribution grid levels (below 69kV). Compared to conventional methods, the proposed method embeds dynamical system models of grid assets in Time-Series (TS) analysis and captures stochastic charging behavior through Monte-Carlo simulation, promising more robust and accurate assessment. Under the proposed method, the Total Cost of Ownership (TCO) of grid assets formulation is re-established. The results of this paper will enable utilities to quantify the capital and operation cost of grid assets induced under various PEV's penetration level and during any time span of interest.  
\end{abstract}

\section{Introduction}

Since their introduction in the 2000s, Plug-in  Electic Vehicles (PEV) have become increasingly popular. According to the International Energy Agency, new registrations of PEV increased by 70\% between 2014 and 2015 \cite{IEAEVOutlook2016}. The power required to charge PEVs is provided at the distribution and potentially sub-transmission level (below 69kV) of the grid. Compared to conventional electrical loads on these levels, PEV loads present distinct characteristics. Firstly, PEVs consume much higher power. As Table~\ref{tab:cstand} shows, at DC Level III, it is possible to charge a 25 kWh battery pack in 10 minutes, which far exceeds the peak power demand for an average household in the U.S. Secondly, the power electronics of PEVs can ramp to full charging level almost instantaneously. For example, it only takes 7 seconds for a 2016 Ford Focus Electric to reach its full charging level after connecting to the grid.

\begin{table}[h!]
\vspace{-0.5em}
\caption{SAE Charging Ratings \cite{sae2011sae}}
\begin{center}
\begin{tabular}{ |p{0.1cm}|p{1.2cm}|p{1.4cm}|p{1.4cm}| } 
\hline
\multicolumn{2}{|c|}{Charging Type} & Voltage (V) & Power (kW) \\
\hline
\multirow{2}{*}{AC} &Level I& 120& up to 2 \\
\cline{2-4}
 &Level II& 240& up to 19.2 \\
\hline
\multirow{3}{*}{DC} &Level I& 200 - 450& up to 36 \\
\cline{2-4}
 &Level II& 200 - 450& 90 \\
\cline{2-4}
 &Level III& 200 - 600& 240\\
\hline
\multicolumn{4}{c}{*DC Fast Level III is not yet finalized.}\\
\end{tabular}
\end{center}
\vspace{-2em}
\label{tab:cstand}
\end{table}

A great amount of work (summarized in \cite{shareef2016review}) have shown that these high density impulsive loads will induce distinguishable effects on the power grid, including fast varying voltage profiles along the distribution feeder and overloading of service transformers. This will consequently affect the grid asset operating status and induce depreciation over the long term. With increasing PEV penetration and improving fast charging technologies, it is critical for utilities to quantify the impact of PEV loads on grid assets and plan for equipment replacement and infrastructure expansion accordingly, in order to ensure service reliability.

Existing studies evaluating PEV's impact fall in two categories: static analysis and Time-Series (TS) analysis. In static analysis, the stress imposed on grid assets is evaluated under the total load composed of the baseline and PEV loads. Most of static analysis results in the worst-case evaluation, which considers the maximum PEV loads induced from coincidental charging \cite{fernandez2011assessment,shafiee2013investigating}. Other work considers the probabilistic distribution of PEV loads connected to the grid and quantifies the induced stress (e.g., power delivery losses and power quality degradation) in the form of probabilistic distribution functions \cite{leou2014stochastic,qian2011modeling}. Static analysis assumes constant grid configurations, so it cannot capture the effect induced by topological change or spontaneous events. For the same reason, it cannot be used to evaluate the loads' impact on discrete operating devices (e.g., online voltage regulators and transformer's secondary tap changers), whose operating status depends on the previous instant. Another major deficiency of static analysis is that it cannot reveal the temporal correlation between grid asset response and stochastic PEV charging activities. These deficiencies can be alleviated in TS analysis.

TS analysis simulates the grid's response over time to one or sets of predetermined load profiles. A few studies adopt TS analysis in PEV's impact evaluation, under deterministic or stochastic settings \cite{shafiee2013investigating,qian2011modeling}. However, the results of these studies do not naturally fulfill utilities' needs of quantifying the long-term cost induced by PEV penetration. This is because (i) TS analysis is simulation-based in essence and case dependent, and (ii) TS analysis only shows the electrical response (i.e.,voltage and current), but grid asset depreciation could depend on response in other dimensions (e.g., winding temperature). For this reason, we propose a method that enables utilities to flexibly evaluate the long-term PEV-induced cost on grid assets. The main contributions of this paper are 
\begin{itemize}
\item Analytically modeling the impact of PEVs' dynamic charging characteristics on temporal response and depreciation of distribution grid assets.
\item Re-establishing the formulation of Total Cost of Ownership (TCO). Original TCO formulation only provides annual average cost of grid assets. The improved formulation enables evaluation of cost rate at any instant or total cost during any time span of interest.
\item Integrating the analytic models with TS analysis and improving the robustness of evaluation with Monte-Carlo simulation under stochastic PEV charging patterns and State of Charge (SoC).
\end{itemize}

This paper assumes the grid operates in steady-state. The dynamical response of grid assets is defined as inter-temporal state change. For example, insulation temperature change and device switching status change. This paper does not address the transient response (i.e. power quality issues) and instability induced from PEV charging. In the paper, ``grid assets'' and ``power delivery equipment'' are used interchangeably. 

The rest of this paper is organized as follows. Section \Romannum{2} presents an overview of the integrated assessment approach. Section \Romannum{3} investigates the in-depth analytic formulation of the approach. The case study and numerical results are shown in Section \Romannum{4}. Finally, Section \Romannum{5} discusses conclusions.

\section{Background}
\subsection{Total Cost of Ownership Analysis}
Based on the way that grid assets depreciate, they can be classified into two categories: continuous loading equipment and discrete operating equipment. The former's long-term cost depends on their thermal loading, while the latter's depends on the devices' operating times. Examples are transformers, which depreciate faster under heavy loading, and voltage regulators, which exhaust after operating for a certain number of times.

TCO analysis is commonly adopted by utilities to assess the long-term cost, comprised of operating cost and capital depreciation, of power delivery equipment. The TCO of discrete operating equipment is conventionally evaluated independent of loading conditions. For continuous loading equipment, its TCO is exemplified by a transformer and expressed as \eqref{eq:TCO} \cite{rural2016guide}. It can be seen that \eqref{eq:TCO} only considers the average annual loading on the equipment. Therefore, the accuracy of long-term cost evaluated with the conventional TCO formulation is highly questionable under PEV of stochastic charging and impulse load characteristics. 
\begin{align}\label{eq:TCO}
\text{TCO (\$/yr)} = C_o + CL\cdot A + LL \cdot B
\end{align}
where $C_o$ is the annual capital cost of the transformer, the rest of the terms are operating cost. $CL,LL$ are transformer core loss and load loss provided by manufacturers, $A$ and $B$ are core loss and load loss factor.
\begin{align}
    &A = N\cdot PEC \\
    &B = LoF \cdot PEC \cdot \hat{P}^2
\end{align}
where $N=8760$ is the total hours in a year,  $\hat{P}$ is the normalized peak loading $\hat{s}/s_R$, 
$PEC$ is the present yearly value of energy cost ([\$/kWh]),which depends on the transformer's insulation life $T_{ins}$, interest rate $i$, and energy cost $EC$ .
\begin{equation}\label{eq:pec}
PEC = EC \cdot \frac{(1+i)^{T_{ins}}-1}{i(1+i)^{T_{ins}}}
\end{equation}
and $LoF$ is transformer loss factor depending on the annual average loading of the transformer $s_{avg}$.
\begin{align}\label{eq:lof}
    LoF = \gamma \frac{s_{avg}}{\hat{s}}+(1-\gamma)(\frac{s_{avg}}{\hat{s}})^2
\end{align}
where $\gamma$ is the dynamic load factor constant.

\subsection{Proposed Method}
\begin{figure}[t]
\centering
\includegraphics[width=0.4\textwidth,height=6cm]{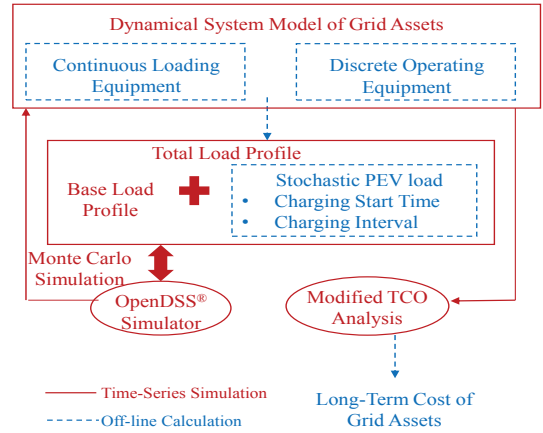}
\vspace{-0em}
\caption{Technical outline of the proposed method.}
\vspace{-0.5em}
\label{fig:diagram}
\end{figure}

To accurately evaluate PEV's impact on the long-term cost of grid assets, we propose an integrated method, which is outlined in Fig.~\ref{fig:diagram} and summarized below. 

\noindent
\emph{Step 1:} Develop analytic models of grid assets.

\noindent
\emph{Step 2:} Evaluate temporal response of grid assets using TS analysis. The stochastic PEV charging activities are captured with Monte-Carlo simulation. The output of this step is comprised of capital cost depreciation and operating cost due to power losses. 

\noindent
\emph{Step 3:} Evaluate the time-varying cost with re-established TCO formulation. The final outputs include cost rate at specific instant $t$ and TCO of grid assets during any time span of interest.
\section{Analytic Models and Formulations}

In this section, we develop the analytic models of grid assets, present the techniques in TS analysis, and re-establish the TCO formulation. 
\subsection{Dynamical System Model of Grid Assets}
Models of grid assets under the same category take similar forms. Due to limited space, we present the dynamical models of transformers and voltage regulators to represent the continuous loading equipment and discrete operating equipment, respectively. These models are adopted to assess the equipment's temporal response in the proposed method. We also derive their Loss of Life (LoL) models.  
\subsubsection{LoL Model of Transformer}
The distribution transformer's lifetime depends on the internal winding hot-spot temperature $Q_{HST}$, which is directly related to loading level $s(t)$ at each instant \cite{international2005iec}. The core of this thermal model has the general form in terms of continuous time differential equations:
\begin{eqnarray} \label{eq:qto}
  &\dot{Q}_{TO}(t)=f_1(\mathbb{E}^2[K(t)],Q_{TO}(t))\\ 
    &\ddot{Q}_{H}(t)=f_2(\mathbb{E}^y[K(t)],\dot{Q}_{H}(t))\\ \label{eq:qht}
    &Q_{HST}(t)=Q_{TO}(t)+\tau_H \cdot \dot{Q}_{H}(t) \label{eq:qhst}
\end{eqnarray}
where $Q_{TO}$ is the top-oil temperature, $\mathbb{E}[K(t)]$ is the expectation of load factor $K(t)=s(t)/s_{R}$ (rated) at each instant obtained from distribution power flow analysis embedded with stochastic methodology, $\dot{Q}_{H}$ is the hot-spot temperature dynamic over top-oil, $\tau_H$ is the hot-spot temperature time constant, and $y$ is the winding exponent power. The compact form of the dynamical system model of \eqref{eq:qto}-\eqref{eq:qhst} can be written  as a stochastic function of continuous loading level.
\begin{align}
    &\dot{Q}_{X}=f(Q_X,s(t)|\mu, \sigma)\\
    &Q_{HST}=a^T \cdot Q_X
\end{align}
where $Q_X=[Q_{TO}~\dot{Q}_{H}]$ and $a=[1 ~\tau_H]^T$.

Then, the actual loss of life $L_T$ for transformer during any time span $[t_1,t_2]$ is derived as \eqref{eq:lt}. The transformer's lifetime $T_x$ can be founded by solving $L_x(0,T_x)=1$. 
\begin{align}\label{eq:lt}
L_x (t_1,t_2)=\frac{1}{T_{ins}} \int \nolimits_{t_1}^{t_2} F_{AA}(t)dt
\end{align}
where $T_{ins}$ is the normal insulation life of the transformer and $F_{AA}$ is the accelerated aging factor defined in \eqref{eq:faa} \cite{board1995ieee}. When $F_{AA}(t)>1$, the lifetime of the transformer is shortened at instant $t$.
\begin{align}
    &F_{AA}(Q_{HST})=exp(\alpha-\frac{\beta}{Q_{HST}(t)+\Omega}) \label{eq:faa}
\end{align}
where $\alpha,~\beta$ and $\Omega$ are design constants of the transformer.

\subsubsection{LoL Model of Voltage Regulator}
Voltage regulators (VR) are essentially a type of tap changing transformer. In the distribution level of power grid, VR are used to regulate voltage deviation from predetermined values. Impulse loads, like PEV, tend to cause fast time-varying and salient voltage deviation, which may result in more frequent operation of VR. VR's lifetimes are determined by their mechanical durability and specified as the total number of operating times. The operation policy of VR can be expressed as \eqref{eq:torus}.
\begin{equation}
h(n)= 
 \begin{cases}
  (V(n)-V_{R})\cdot \frac{1}{\kappa}, \text{ if } V(n)\in[h_{min},\underline{\epsilon}]\cup[\bar{\epsilon},h_{max}]\\
    h_{max}, \qquad \text{if } h(n-1)+\Delta h(n)\geq h_{max}\\
    h_{min}, \qquad \text{if } h(n-1)-\Delta h(n)\leq h_{min}\\
  \end{cases} \label{eq:torus}
\end{equation}
where $h(n)$ is the VR tap position at the $n^{th}$ sampled instant after each operating cycle, $V(n)$ is the discrete voltage level calculated from power flow, $V_R$ is the regulated voltage, $\kappa$ is the VR step-size, $[\underline{\epsilon},\bar{\epsilon}]$ is VR's dead-band, and $h_{max}$, $h_{min}$ are maximum and minimum tap position.

By observing such change of tap positions triggered by voltage variation, the LoL of VR during any time span $[n_1,n_2]$ can be obtained in \eqref{eq:lvr}, and the VR's lifetime $T_v$ can be founded by solving $L_{V}(0,T_v)=1$.
\begin{align}\label{eq:lvr}
L_{V}(n_1,n_2)=\frac{1}{N_{op}}\sum_{n_1}^{n_2} |h(n)-h(n-1)| 
\end{align}
where $N_{op}$ is the VR's empirical maximum number of tap operations.
\subsection{Time-Series Analysis considering Stochastic PEV Charging Patterns}
To truly reflect stochastic PEV charging patterns, Monte-Carlo Simulation (MCS) is implemented in Time-Series (TS) analysis. The procedure is detailed in the sequel. First, the stochastic PEV charging activities are modeled with two random variables: the charging start time $t_{s}$ and the charging interval $\Delta t$. The latter is an explicit function of initial State of Charge (SoC) of individual PEV battery at the beginning of each charging period, given battery capacity $C$ ([kWh]) and charging power level $P^{PEV}$ ([kW]), i.e., $\Delta t=C\cdot(1-SoC)\big{/}P^{PEV}$.

Second, the TS load profiles for each node are generated according to the probability distribution of PEV charging patterns. The total load at node $h$ is the sum of the baseline load and PEV load,  $P_h(t)=P_h^o (t)+P_h^{PEV}(t)$.

Thirdly, the power flows in the grid are simulated in multiple iterations under generated TS load profiles and are input into the analytic models of grid asset response simultaneously. The law of large numbers indicates that as sample size gets large enough, the expected value of model outputs can be approximated by taking the sample mean of the MCS output results. The final outputs are TS probability distribution of grid (assets) response at every node. 
\subsection{Modified TCO Analysis}
The outputs of TS analysis enable us to accurately assess the LoL of power delivery equipment in the grid with PEV loads during any time span of interest. Therefore, the long-term cost of the equipment can be obtained by re-establishing the TCO formulation. For VR, the TCO can be simply expressed as: 
\begin{equation}\label{eq:mtcovr}
\text{TCO}(n_1,n_2)=L_{V}(n_1,n_2)\cdot C_0
\end{equation}
where $L_{V}(n_1,n_2)$ is specified in \eqref{eq:lvr} and $C_0$ is the VR's capital cost. 
For transformers, the TCO can be formulated as

\begin{align}\label{eq:mtcot}
\text{TCO} (t_1,t_2) = &L_x (t_1,t_2)\cdot C_o \\ \nonumber
&+ CL\cdot A(t_1,t_2) + LL \cdot B(s,t_1,t_2)
\end{align}
where $L_T(t_1,t_2)$ is specified in \eqref{eq:lt} and other parameters are specified in Section II.A. $PEC$ in \eqref{eq:pec} is modified to reflect the future cost in $[t_1,t_2]$ to the present day value as 
\begin{equation}
PEC = \frac{EC}{i}[\frac{1}{(1+i)^{t_1}}-\frac{1}{(1+i)^{t_2}}]
\end{equation}
and $LoF$ in \eqref{eq:lof} is modified to capture time-varying loading level under stochastic PEV charging patterns as
\begin{align}
    LoF(s,t) = \gamma \frac{\mathbb{E}[s(t)]}{\hat{s}}+(1-\gamma)(\frac{\mathbb{E}[s(t)]}{\hat{s}})^2 
\end{align}

In both \eqref{eq:mtcovr} and \eqref{eq:mtcot}, the first term reflects capital cost of the equipment due to the accelerated depreciation resulted from extra stress imposed by PEV loads, while the other terms in \eqref{eq:mtcot} reflects the operating cost induced from stochastic TS load profiles.  
\section{Demonstration of Proposed Method}

The proposed method is demonstrated on the standard IEEE 13 node test feeder with MATLAB and OpenDSS. The test system has peak base load (the load without PEV) of $\hat{P^o}\approx 3000~kW$. To observe progressive impact, multiple PEV penetration level ($PL$) scenarios from $0$ (no PEV) to $300\%$ in steps of $50\%$ are investigated. The upper bound of $PL$ is set based on the scenario that every household owns a PEV, which is predicted realistic in future \cite{singer2016consumer}. $PL$ is defined as the aggregated charging power capacity of PEV throughout the system divided by the peak annual total base load. 
\begin{align}
    PL=(\sum_{h=1} P_h^{PEV} \big{/} \hat{P^o}) \times 100%
\end{align}

As variable types and charging patterns of PEV are available in the current market, it is imperative to have a \textit{ceteris paribus} case in our study. Hence, it is assumed that the same type of $C=23~kWh$ battery package is used for all PEV loads. Each PEV is charged at most once per day. Moreover, level \Romannum{2} charging infrastructure ensures the power demand per charge is uniform and constant as $P^{PEV}=10~kW$ in order to accommodate the trend of public fast charging technology (see Table \Romannum{1}). According to historical PEV driving survey data \cite{leou2014stochastic}, the probability distribution of these two sets of random variables can be assumed to follow Gaussian distribution as $t_s\sim \mathcal{N}~(20.5,4.5^2)$ and $\Delta t \sim \mathcal{N}~(1.2,0.6^2)$. In addition, it is assumed that the charging start time and initial SoC are independent for each individual PEV. Since the Gaussian distribution has tail parts which produce unreasonable values, the random generation process for charging interval $\Delta t$ has zeroed-out the negative value scenarios. Finally, the MCS samples $100$ scenarios of $PL$ in TS analysis.  

\subsection{Grid Asset Depreciation Analysis}
\begin{figure}[t]
\centering
\begin{minipage}{0.5\textwidth}
\subfloat[Transformer Daily Aging Factor]
{  
  \includegraphics[width=\linewidth, height=3cm]{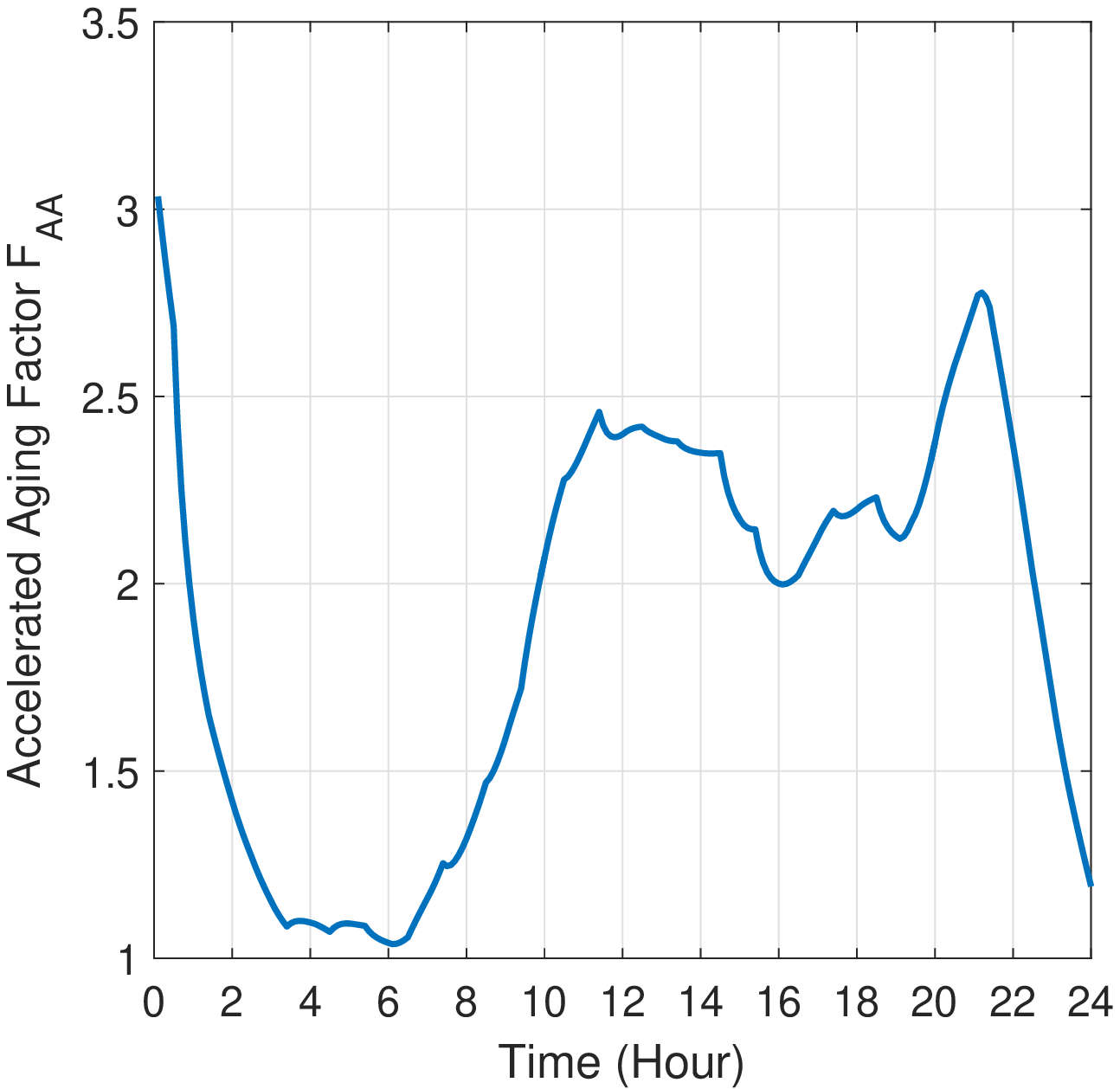}
  \label{voltage violation}
}
\\
\vspace{-0.2em}
\subfloat[VR Daily Tap Operation]
{
  \includegraphics[width=\linewidth, height=3cm]{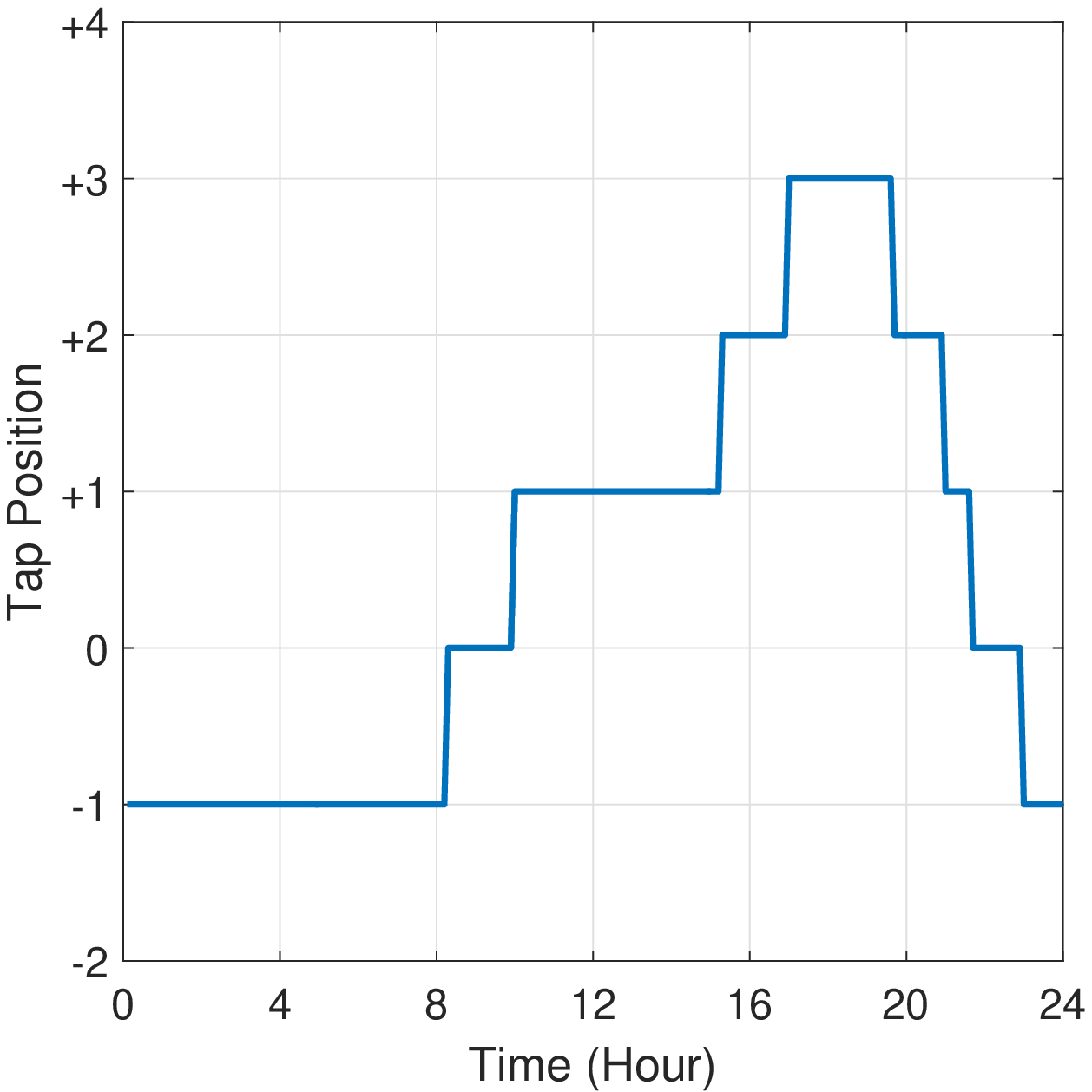}  
  \label{probability}
}
\vspace{0.3em}
\caption{Dynamic Asset Depreciation for $300\%$ $PL$}
\label{voltage violation and probability}
\end{minipage}
\end{figure}

\begin{figure}[b]
\vspace{-1.3em}
\centering
\includegraphics[width=0.48\textwidth,height=6cm]{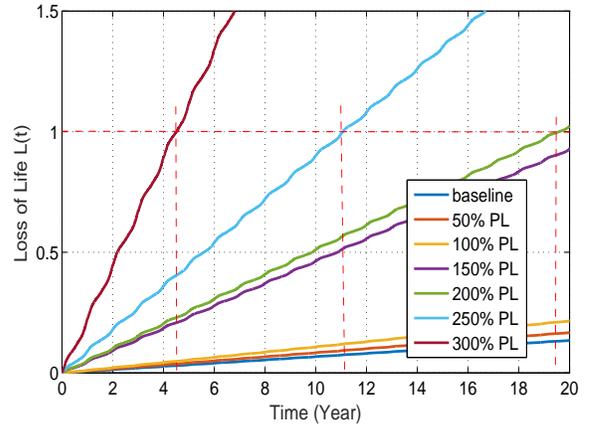}
\vspace{0em}
\caption{Transformer Loss of Life under various PEV penetration levels.}
\vspace{-0.5em}
\label{transformer}
\end{figure}

\begin{figure}[t]
\vspace{-0em}
\centering
\includegraphics[width=0.48\textwidth,height=6cm]{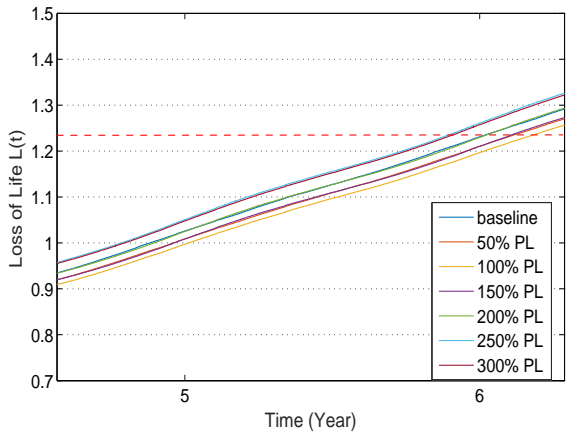}
\vspace{-0em}
\caption{Voltage Regulator Loss of Life under various PEV penetration levels.}
\vspace{0em}
\label{vr}
\end{figure}

The accelerated aging factor $F_{AA}$ and VR tap operation pattern in a typical day (0.1-hour resolution) for $300\%$ $PL$ is shown in Fig.~\ref{voltage violation and probability}. The design parameters of the transformer in \eqref{eq:qhst}, \eqref{eq:faa} are from \cite{kuss2011application}. It can be observed that the temporal response and depreciation of the selected transformer and VR is strongly correlated with PEV charging activities.
The equipment LoL $L(t)$ until instant $t$ under various $PL$ is shown in Fig.~\ref{transformer} and Fig.~\ref{vr}. The equipment lifetime $T_x$ and $T_v$ intersect with $L_x(T_x)=1$ and $L_v(T_v)=1$. It is observed that the transformer's lifetime is greatly shortened with increasing $PL$. On the other hand, the VR's lifetime is not necessarily correlated with increasing $PL$. This is due to the tap position limits and operating cycle settings of VR. Even though a higher $PL$ is likely to cause greater voltage deviation and more salient TS load profiles, the tap of VR will not operate if it has reached the limit and will not operate more frequently than the operating cycle.

\subsection{Long-term Cost Evaluation of Grid Assets}
\begin{figure}[ht]
\vskip-1.2ex
\centering
\includegraphics[width=0.48\textwidth,height=6cm]{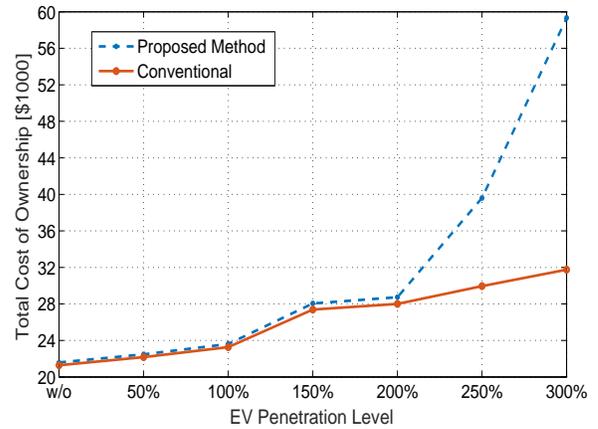}
\vspace{0.2em}
\caption{Long-term cost of transformer estimated with the proposed method and conventional TCO formulation}
\vspace{-0.5em}
\label{tco}
\end{figure}

The long-term cost of a transformer is estimated with the proposed method and compared with conventional TCO formulation described in Section II.A. The results are compared over the transformer's insulation lifetime $T_{ins}$. If the transformer is exhausted at $T_x$ before $T_{ins}$ due to the extra stress imposed by PEV loads, then a new transformer is purchased and its induced cost (comprised of capital cost and operating cost) is added to the total cost. The parameters of the assessed service transformer are obtained from an anonymous vendor and summarized in Table \Romannum{2}. The long-term costs estimated with two methods are plotted in Fig.~\ref{tco}.  
\begin{table}[h!]
\centering
\caption{TCO Parameters and Specifications}
 \begin{tabular}{||c|c||} 
 \hline
 Parameters & Value \\ 
 \hline\hline
 $|s_R|$ [kVA] & 500  \\
 
  $CL$ [W] & 960  \\
 
 $LL$ [W] & 5100  \\
 
  $EC$ [\$/kWh] & 0.05  \\
 \hline
  $\gamma$  & 0.2  \\
 
  $i$ [\%] & 5  \\
 
  $C_o$ [\$] & 4575  \\ 
 
  Evaluation Period [yr] & 20  \\ 
 \hline
\end{tabular}
\end{table}
As shown in Fig.~\ref{tco}, the results of both methods indicate that the long-term cost of the transformer is greatly increased with $PL$. Moreover, results are very close at low $PL$, and the difference becomes noticeable when $PL$ reaches $150\%$. When $PL$ is greater than $200\%$, the proposed method assesses much higher long-term cost than conventional TCO. This difference in trend is attributed to the fact that service transformers are usually over-sized for reliability concerns. Therefore, a relative low $PL$ is not likely to cause noticeable adverse impact on transformer operation. However, when the grid hosts more PEV, the impact can only be captured with the proposed method. 

A further question to ask is whether it is more reasonable to use a larger size transformer under high $PL$, which will essentially bring the results of the two methods to the same values. The answer could be case dependent. For example, sometimes a larger transformer could cost more than replacing a small transformer after its end of life, while other times the reverse is true. Nevertheless, even if the planning strategy might conceal the inaccuracy of the conventional TCO method, the fidelity of the proposed method is demonstrated at every $PL$. Moreover, the proposed method enables evaluation of equipment long-term cost over any time span of interest, which provides great flexibility to utilities planning work.

\section{Conclusion}

With increasing PEV penetration and improving fast charging technologies, it is critical for utilities to quantify the impact of PEV loads on grid assets and plan for equipment replacement and infrastructure expansion accordingly to ensure service reliability. The impulsive and high-power-density characteristics of PEV loads make conventional assessment methods of load impact unsuitable. To address this challenge, this paper proposes an integrated method to investigate the long-term impact of PEV charging on temporal response and depreciation of power delivery assets in sub-transmission and distribution grids. The main contributions of this paper include (i) developing analytic models of grid assets depreciation under time-varying load profiles; (ii) Re-establishing the formulation of Total Cost of Ownership (TCO), which enables flexible evaluation of long-term cost of equipment during any time span of interest; and (iii) improving the robustness of evaluation under stochastic PEV's charging patterns by integrating Monte-Carlo simulation in time-series analysis. The results of this paper can be developed software planning tools for utilities. The fidelity of the proposed method is demonstrated on the IEEE 13-Node test feeder.

\bibliographystyle{IEEEtran}
\bibliography{reference}

\end{document}